# RANS SIMULATIONS OF TURBULENT ROUND JETS IN THE PRESENCE OF DENSITY DIFFERENCE AND COMPARISON WITH HIGH-RESOLUTION EXPERIMENTAL DATA

**Jiaxin Mao, Sunming Qin, Victor Petrov and Annalisa Manera**

Department of Nuclear Engineering and Radiological Sciences, University of Michigan – Ann Arbor, 2355 Bonisteel Blvd, Ann Arbor, MI 48109, USA

**ABSTRACT**

Turbulent round free jets, as one of the canonical flow configurations, have been investigated since the beginning of the last century. The predictive capabilities of Reynolds-averaged Navier–Stokes (RANS) models for these flows have also been studied in detail. The behavior of turbulent round free jets in the presence of density difference is however still not fully understood. For these types of flows, even standard Large Eddy Simulation (LES) models are not suitable because the assumption of turbulence isotropy at the smaller eddy scales breaks down in the presence of stratified layers. Recently, high-resolution experiments have been performed for turbulent jets in the presence of density difference [1].

In this paper, the novel experimental data reported by Qin et al. [1] are used to assess the predictive capability of the Realizable k-ε (RKE) model and Reynolds stress transport (RST) model for buoyant jets and understand the reasons for discrepancies. In particular, we present the comparison between simulation results of a turbulent buoyant jet flow in the self-similar region with high-resolution experimental data obtained for a jet injected from a 2 mm nozzle into a 300×300×300 mm$^3$ tank, with nominal Reynolds number equal to 10,000. Results show that streamwise velocity profiles predicted by the RST model had good agreement with experimental data, while the larger spreading rate was predicted by the RKE model. For turbulent statistics, turbulent kinetic energy witnessed a discrepancy in the center region, with shear stress well predicted for both models. Comparison of the turbulent kinetic energy production term with experimental data revealed reasons for the discrepancy and also showed that the gradient of the streamwise velocity in the crosswise direction contributes the most to the turbulent kinetic energy production. Investigation of model coefficients of the turbulent dissipation equation for the RKE model has revealed that $C_{\varepsilon 2}$ is critical in model accuracy.

## 1. INTRODUCTION

Turbulent free jets, characterized by momentum transport and turbulent mixing, have wide engineering applications. They are relevant for reactor applications when considering steam jets in the containment following a SBLOCA, emergency core coolant injections during accident scenarios, pressurized thermal shock in PWR reactor vessels, hot jets at the core outlet of sodium-cooled reactors interacting with the reactor pool upper plenum, etc. The first study of turbulent jets dates back to the time when Tollmien [2] first theoretically investigated circular jets and since then the turbulent jet flow has been widely studied. Turbulent round jet, as one of the canonical turbulent free-jet flow configurations, is one of the classical candidates for studying turbulent flow thanks to its self-similarity feature and axisymmetric configuration. A thorough literature review of research efforts on this type of flow has been done by Ball et al. [3]

Though its physics has long been known to be governed by Navier-Stokes (NS) equations, analytical solutions exist only for limited cases due to the non-linear property of these equations. This leads to the fact that computational fluid dynamics (CFD) remains an indispensable tool in understanding the details in addition to experiments. In regard to numerical approaches, direct numerical simulation (DNS) and large eddy simulation (LES) which solves space-averaged Navier-Stokes equations, though have been used by different researchers to simulate turbulent round jets ([4], [5], [6], [7], [8], [9]), are still not commonly employed due to its high demand on computational power. As a compromise between fidelity and efficiency, Reynolds averaged Navier-Stokes (RANS) approach has been prevailing as a tool to gain insight in studying problems involving turbulent jets.

Based on Reynolds decomposition which decomposes time-dependent variables into time-averaged mean value and its instantaneous fluctuation [10], Reynolds averaged Navier-Stokes (RANS) approach solves mean values of time-dependent variables. The time-averaged NS equations, in the format of Einstein notation, are presented below, where ρ is density, Φ is mean scalar with its fluctuation represented by φ′ and U is mean velocity with its fluctuation u′.

$$\frac{\partial \rho}{\partial t} + \frac{\partial \rho U_j}{\partial x_j} = 0 \tag{1}$$

$$\frac{\partial \rho U_j}{\partial t} + \frac{\partial \rho U_i U_j}{\partial x_j} = -\frac{\partial P}{\partial x_i} + \frac{\partial}{\partial x_j}\left(\mu\left(\frac{\partial U_i}{\partial x_j} + \frac{\partial U_j}{\partial x_i}\right) - \frac{2}{3}\mu\frac{\partial U_k}{\partial x_k}\delta_{ij} - \rho\overline{u_i'u_j'}\right) + g_i(\rho - \rho_o) \tag{2}$$

$$\frac{\partial \rho \Phi}{\partial t} + \frac{\partial \rho \Phi U_j}{\partial x_j} = \frac{\partial}{\partial x_j}\left(\Gamma\frac{\partial \Phi}{\partial x_j}\right) - \frac{\partial}{\partial x_j}\left(\rho\overline{u_j'\varphi'}\right) \tag{3}$$

As can be seen from the Reynolds equations, turbulence modeling is required to find closure for the Reynolds stresses $\rho\overline{u_i'u_j'}$ in the momentum equation and the turbulent fluxes $\rho\overline{u_j'\varphi'}$ in the scalar transport equation.

Often the turbulent-viscosity hypothesis is used, which assumes the Reynolds stresses to be proportional to the mean rate of strain with a positive scalar coefficient called turbulent viscosity. This approximation is used in several so-called eddy viscosity models, including the very well known two-equation k-ε model. In Reynolds stress models instead, transport equations for second-moment closure are solved instead. The turbulent scalar flux, both in Reynolds stress models as well as for eddy-viscosity models, are typically assumed to be proportional to the gradient of the mean flow scalar distribution, according to the standard gradient diffusion hypothesis (SGDH). A detailed discussion of turbulence models can be found in Pope [11].

These turbulence models have been extensively used in studying turbulent round free jets in a uniform environment. For example, comparison of simulation results from these turbulence models with experiments in predicting turbulent round jets can be found in ([12], [13], [14], [15]), and work of using those models to do parameter investigation and optimization has been reported in ([16], [17], [18]). Results from their work have shown that the k-ε model, while still overpredicts the spreading rate of round jets, can nevertheless give reasonable predictions compared to other turbulence models.

However, the performance of the RANS model has not been extensively investigated in the presence of density differences, though buoyancy-driven turbulent jets are relevant in reactor safety scenarios for LWRs as well as advanced reactors. The presence of density gradients would cause additional gradient transport effect in momentum and enhancement or suppression of turbulence production. Efforts to take into account buoyant effect in turbulence models has resulted in the implementation of an additional source terms for turbulence production and dissipation ([19], [20], [21], [22]). Nevertheless, the capability of RANS turbulence models to predict turbulent buoyant round jets have not been fully understood due to the lack of detailed experimental data. Recently Qin et al. [1] conducted high-resolution experiments of turbulent buoyant round jets, with detailed measurements of the instantaneous flow field. In this paper, the novel experimental data are used to assess the predictive capability of the Realizable k-ε (RKE) model and Reynolds stress transport (RST) model for turbulent round buoyant jets and identify the reasons for discrepancies. The commercial CFD code STAR-CCM+ v13.04 has been used for the simulations, and the profiles of both first-order and second-order turbulence flow statistics in the self-similar region are compared with high-resolution experimental data.

## 2. EXPERIMENT DESCRIPTION

The experimental facility, miniDESTROJER (mini DEnsity Stratified Turbulent ROund free Jet expERiment) at the University of Michigan has been built and used for the measurement of turbulent buoyant jets flows. In Fig.1 (left) the front view of the experimental facility is shown. A sketch of the round-jet flow in polar-cylindrical coordinates and the associated dimensions are shown in Fig. 1(right). The tank is made of glass with a dimension of 300×300×300 mm$^3$, on the bottom of which a nozzle is present, characterized by a diameter of D=2mm and an inlet region upstream of the tank section of L/D = 25. A synchronized high-speed PIV/PLIF system is used to measure the flow field in the vertical mid-plane of the jet. A density difference with a ratio of 3.16% is achieved by mixing two miscible solutions with different densities. Experiments have been conducted for four cases, with detailed fluid properties shown in Table I. All four cases have the same nominal Reynolds number equal to 10,000.

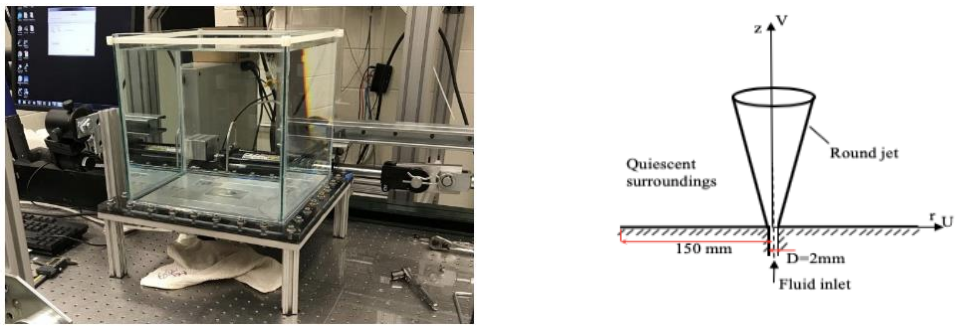

**Figure 1. MiniDESTROJER experimental facility (left) and sketch of jet flow (right)**

**Table I. Fluid Properties test matrix**

| Case | $\dot{m}[kg/s]$ | $\Delta\rho$ | $\rho_{jet}\,[g/cm^3]$ | $\rho_{sur}\,[g/cm^3]$ | $\mu_{jet}\,[Pa\cdot s]$ | $\mu_{sur}\,[Pa\cdot s]$ |
|---|---|---|---|---|---|---|
| **D029** | 0.0190 | 0 | 1.012 | 1.012 | 0.001184 | 0.001184 |
| **D031** | 0.0181 | 0 | 1.044 | 1.044 | 0.001139 | 0.001139 |
| **D033** | 0.0190 | -3.16% | 1.012 | 1.044 | 0.001184 | 0.001139 |
| **D035** | 0.0181 | +3.16% | 1.044 | 1.012 | 0.001139 | 0.001184 |

## 3. CFD MODELING AND SIMULATION

The CFD computational domain consists of a 300×300×250 mm$^3$ box, with an inlet section of 2 mm in diameter extruded by a total length of 5mm. The computational mesh was generated with ANSYS ICEM-CFD and consists of a conformal, structured hexahedral mesh with 13.3 Million cells. Because of the nature of the experiment, the mesh was refined along the jet trajectory. The mesh is fine enough to capture the flow structures found in the PIV measurements. In addition, by taking advantage of the symmetry of the experiment, 2D-axisymmetric simulations were executed as well, to reduce the computational time. The mesh generated for the 2D-axisymmetric simulations was made by only keeping the cells at the mid-plane of the geometry. In this manner, the same cell distribution is maintained between the 3D and 2D simulations (at the mid-plane). Pre-test simulations have shown that the 2D-axisymmetric simulation provides an almost identical profile to that of the 3D simulation. Snapshots of the mesh are shown in Fig. 2 below.

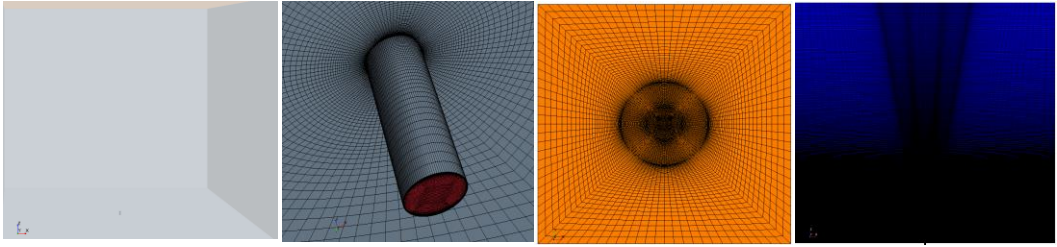

**Figure 2. Geometry and mesh presentation (geometry, inlet, top view, cross-section view)**

It is noticeable that in the experiment, the tank top was higher than the level of tank fluid, which resulted in a slight increase of fluid level due to the accumulation of injected fluid through the jet. Due to the long distance between the area of interest and the tank top, and in view of the relatively small surface level increase (roughly 4 mm in comparison with 250 mm of water level), the influence of the outlet boundary condition on the jet can be assumed to be negligible. Therefore, for each run the jet velocity was specified as inlet boundary condition, the tank walls were set as no-slip boundaries, and the top of the computational domain was set as an outlet boundary condition. Fully-developed flow conditions were imposed at the inlet boundary and numerical schemes were set to be second order. Simulations were performed using the RKE and the RSM turbulence models. While for the RKE model, steady state simulations were sufficient to reach convergence, transient simulations had to be performed when using the RSM model in order to achieve satisfactory convergence of the numerical solution. The transient simulation was stopped once steady-state conditions were achieved. The comparison between simulations and experimental data are discussed in the next section.

## 4. RESULTS AND DISCUSSION

Radial profiles of quantities of interest are compared in the jet self-similar region at four locations, namely z/D=50,60,70,80 from the jet exit. In addition to the mean velocity field and the turbulent statistics, the production term of the turbulence kinetic equation is also compared with experimental data to gain deeper insight in the performance of the turbulence models investigated in this paper.

### 4.1 Streamwise Velocity Profiles

Fig. 3 shows the streamwise velocity contour downstream of the z/D=45 location for the four runs summarized in Table I. From the contour plots in Fig. 3, it can be clearly seen that, as the jet penetrates further through the surrounding fluid, center velocity decreases due to dissipation of momentum and jet spreading. In addition, the RST model predicts larger jet center velocity compared to the RKE model for all four cases, regardless of the density differences. In comparison with experiments, it is clear that the RST model has a better match than the RKE model, especially in the jet center. One noticeable difference between the two uniform cases and the two cases with density differences is that the latter has negative

NURETH-18, Portland, OR, August 18-22, 2019                                                                                           2655

streamwise velocity in the surrounding field. This could come from the enhanced mixing between jets and surroundings.

To better illustrate the accuracy of both models, Fig. 4 below shows the radial streamwise velocity profiles for four cases at four axial locations. It can be seen that overall the RST model has better agreement with experimental data in the prediction of velocity profile, while results from the RKE model show consistent discrepancy compared to experimental data, except for case D033, in which light fluid was injected into the heavy fluid. Simulation by the RKE model underestimates velocity in the jet center region while overestimates velocity in the shear layer region for two reference cases and heavy-to-light case. This indicates that for these three cases, the RKE model has predicted jets with a higher spreading rate and decay speed.

In addition, as z/D increases, downstream from the nozzle, the discrepancy between simulation and experimental results tends to get smaller regardless of the density difference. It is known that moving further away downstream of the nozzle inlet, most momentum has been dissipated and the dissipation rate consequently slows down. Intuitively, one might conclude that the overprediction of the spreading rate arises from the overestimation of turbulence dissipation. In order to check this hypothesis, additional experiments will be performed in the future to measure also the third (out of plane) velocity component.

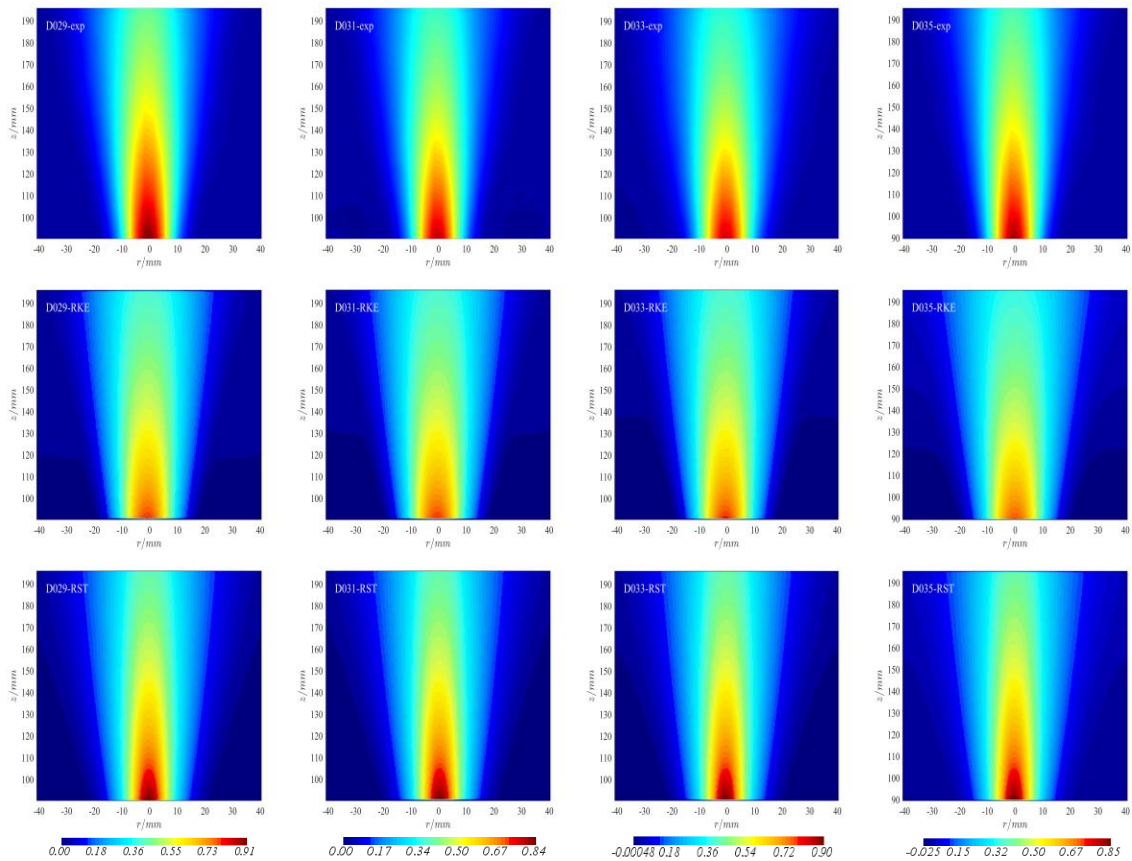

**Figure 3. Streamwise velocity contours (m/s) [top row: exp, middle row: RKE, bottom row: RST]**

To further validate our experimental and simulated results, velocity profiles for case D029 are normalized based on the distance from the jet exit and jet centerline velocity and then compared with similar results found in literature as shown in Fig. 5 below. Fig. 5 (left) shows our normalized results and experimentally-obtained profiles of velocity given by three researchers ([10], [15]). Clearly it can be seen that except the result given by Rodi [10], our experimental profile has a good match with the other two. While the RST model shows good agreement on normalized profile with experiments, the RKE model provides a broader profile off jet center. The trend on predicting broader velocity profile, on another word, also equals to overpredict decay rate based on momentum conservation. Similar simulations have also been conducted by Miltner et al. [15] and Wilcox [10] for uniform jets and their results are shown in Fig. 5 (right). Though profile from the RST model given by Miltner et al. [15] deviates more from experimental data than our result, the same broader profiles are witnessed for the RKE model.

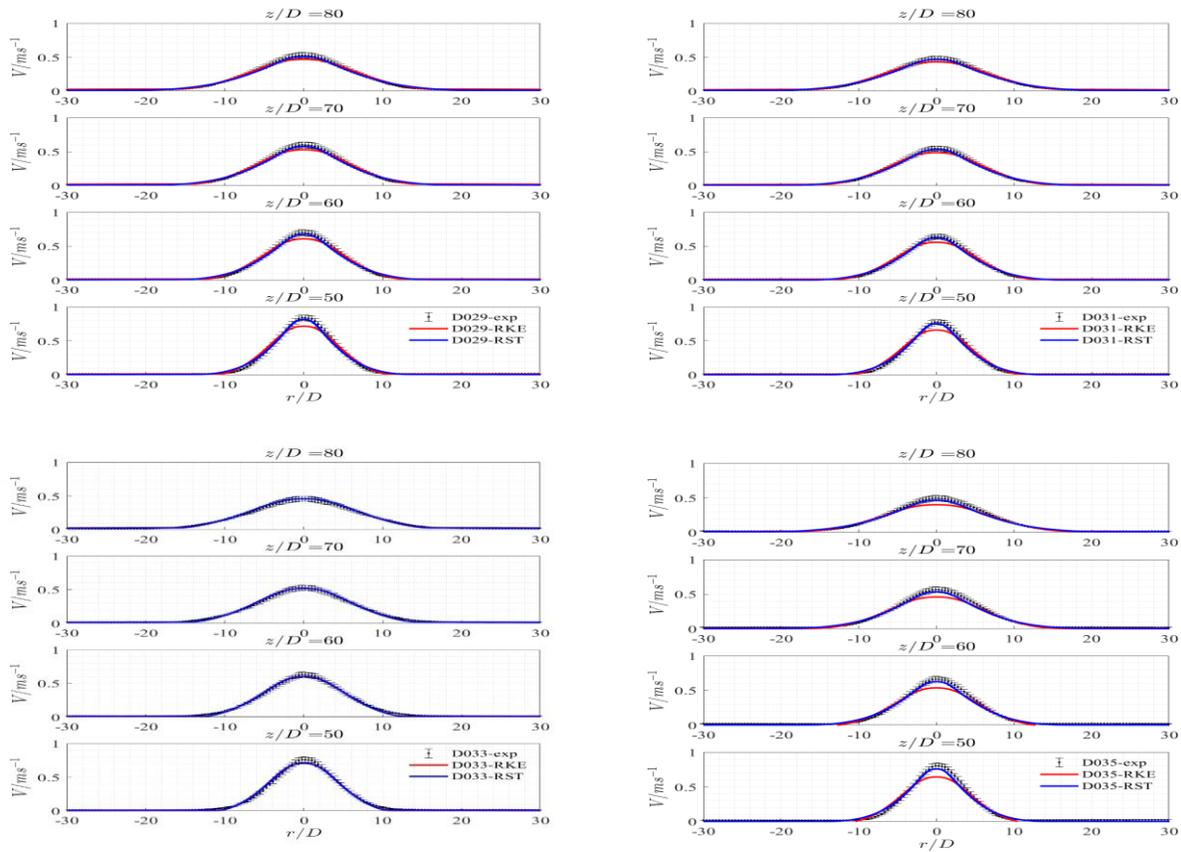

Figure 4. Streamwise velocity profiles

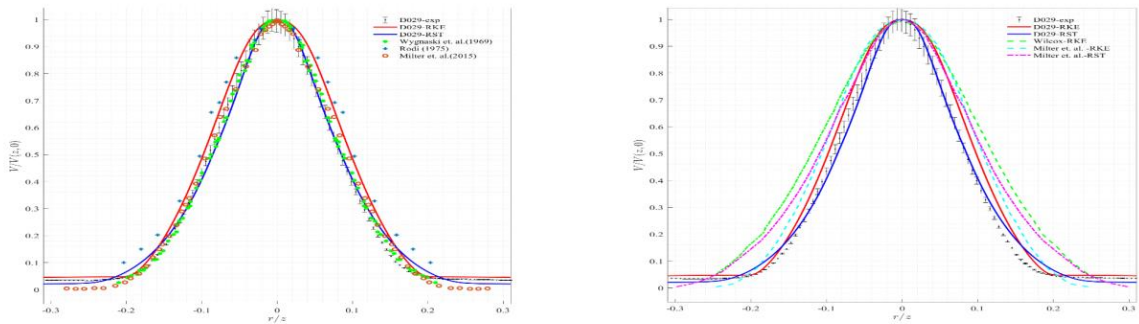

Figure 5. Normalized velocity profiles in comparison with results from literature

## 4.2 Reynolds Stress, Turbulent Kinetic Energy and Turbulent Eddy Viscosity

To investigate the reasons for the discrepancy observed between the experimental data and the RKE model, as shown in Fig. 6, we focus our attention on turbulence parameters. Looking at Eq. (2), it is obviously necessary to have closer scrutiny of the Reynolds stresses. Reynolds stresses are composed of shear stresses ($\overline{u_i' u_j'}$) and normal stresses. The radial profiles of the shear stress obtained for the four runs are compared to experimental data in Fig. 6. It can be seen that the obtained shear stress profiles are in good agreement with the experimental data, even for cases in which density differences are involved. This aligns with the conclusion from the experiments that the turbulence momentum transfer is almost identical regardless of the density effect [1].

As mentioned in section 1, while for the RST model transport equations are solved for the individual Reynolds stresses, in the RKE model the turbulence eddy-viscosity assumption is adopted as closure for the Reynolds stresses, as according to the following formulation:

$$-\overline{u_i' u_j'} = \nu_t \left( \frac{\partial U_i}{\partial x_j} + \frac{\partial U_j}{\partial x_i} \right) - k \delta_{ij} \qquad (4)$$

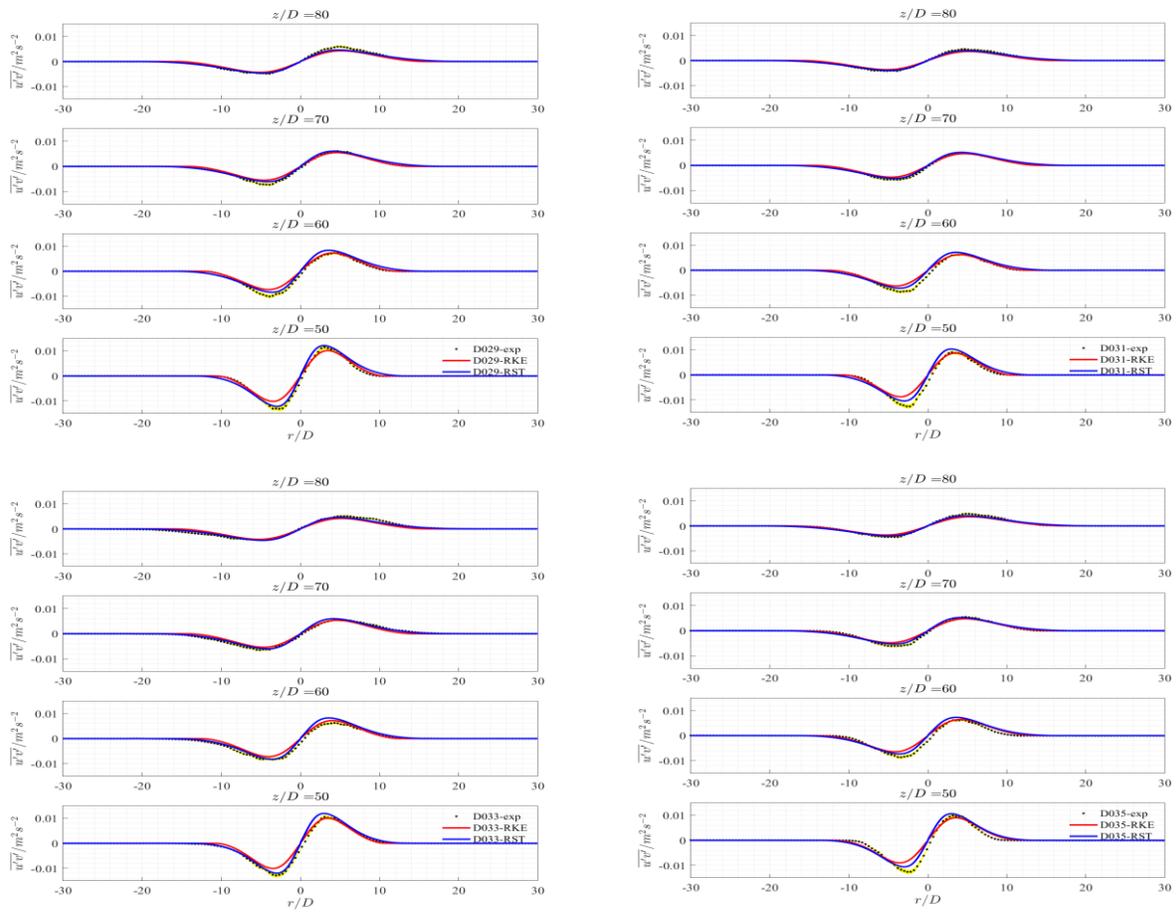

**Figure 6. Shear stress profiles**

where, $v_t$ is the turbulent eddy viscosity and k is the turbulent kinetic energy. Eq. (4) can be used to compute the turbulent eddy viscosity. Based on the Reynold stresses computed by the RKE model and as measured in the experiments, the combination of accurately predicted shear stress profiles and underpredicted streamwise velocity profiles makes it obvious that turbulence eddy viscosity must be overpredicted, as demonstrated in Fig. 7. Similar to the streamwise velocity profiles, the eddy viscosity profiles also deviate in the jet center region, while exhibiting relatively good agreement off jet center.

In the RKE model, two additional transport equations are solved for the turbulent kinetic energy k and the turbulent dissipation rate ε respectively, and the eddy viscosity is given by:

$$v_t = 0.09 \frac{k^2}{\varepsilon} \quad (5)$$

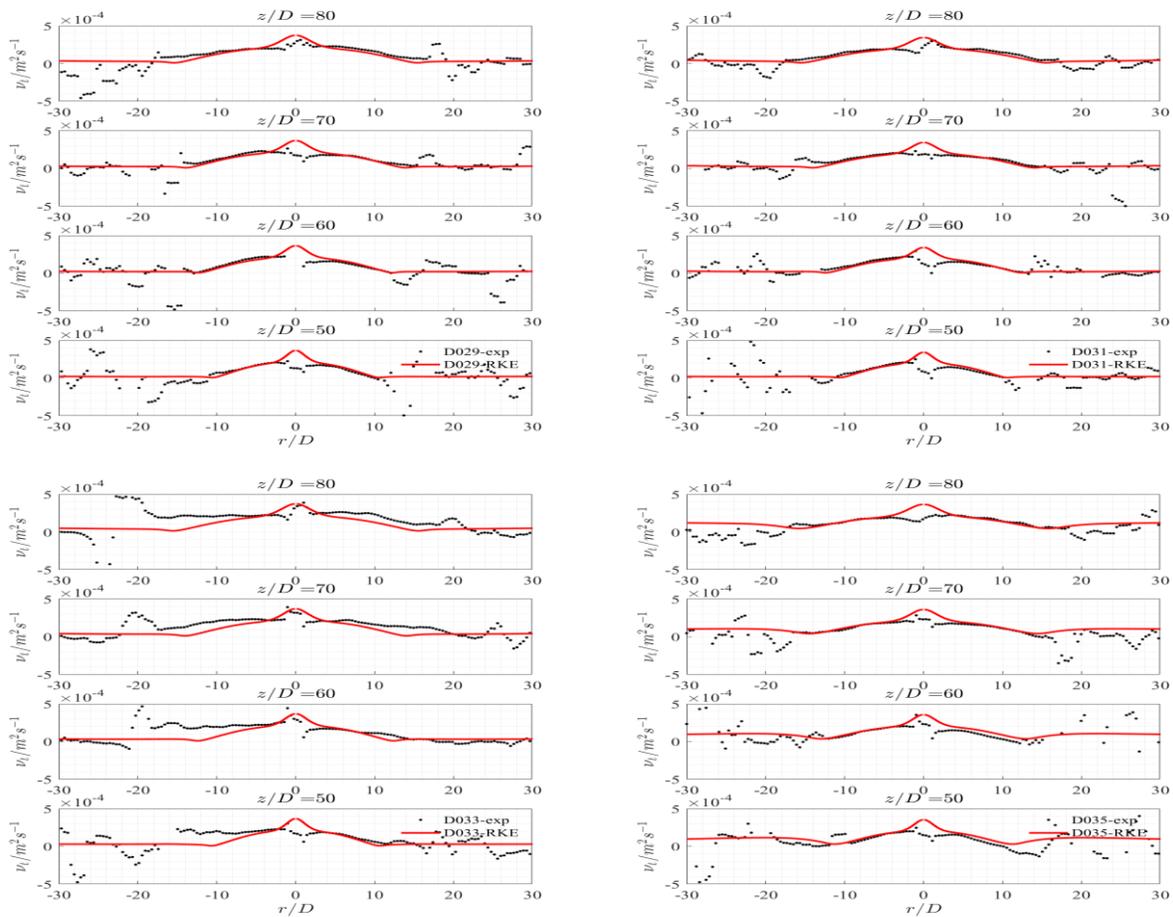

**Figure 7. Turbulent eddy viscosity profiles**

In the current experiments, only two-dimensional data were acquired for the r-z plane. While the turbulence kinetic energy can be extracted from the experimental data, the turbulence dissipation rate is not, being a term that would require measurements of the out-of-plane velocity component as well. Defining the 'pseudo' turbulent kinetic energy in the two-dimensional as:

$$\tilde{k} = \frac{1}{2}(u'u' + v'v') \qquad (6)$$

where $v'v'$ and $u'u'$ are the normal stresses in the streamwise and radial direction respectively, the predicted turbulent kinetic energy for both models is compared to the experimental data in Fig. 8. It can be clearly seen that the RKE model underestimates turbulent kinetic energy in the center region, while exhibits reasonably good agreement in the shear layer. This is different from the results shown by Aziz et al. [14], which predicted lower value in the center and higher value in the shear layer by using the standard k-ε model. The RST model instead tends to overestimate the turbulence kinetic energy, especially at locations closer to the jet inlet.

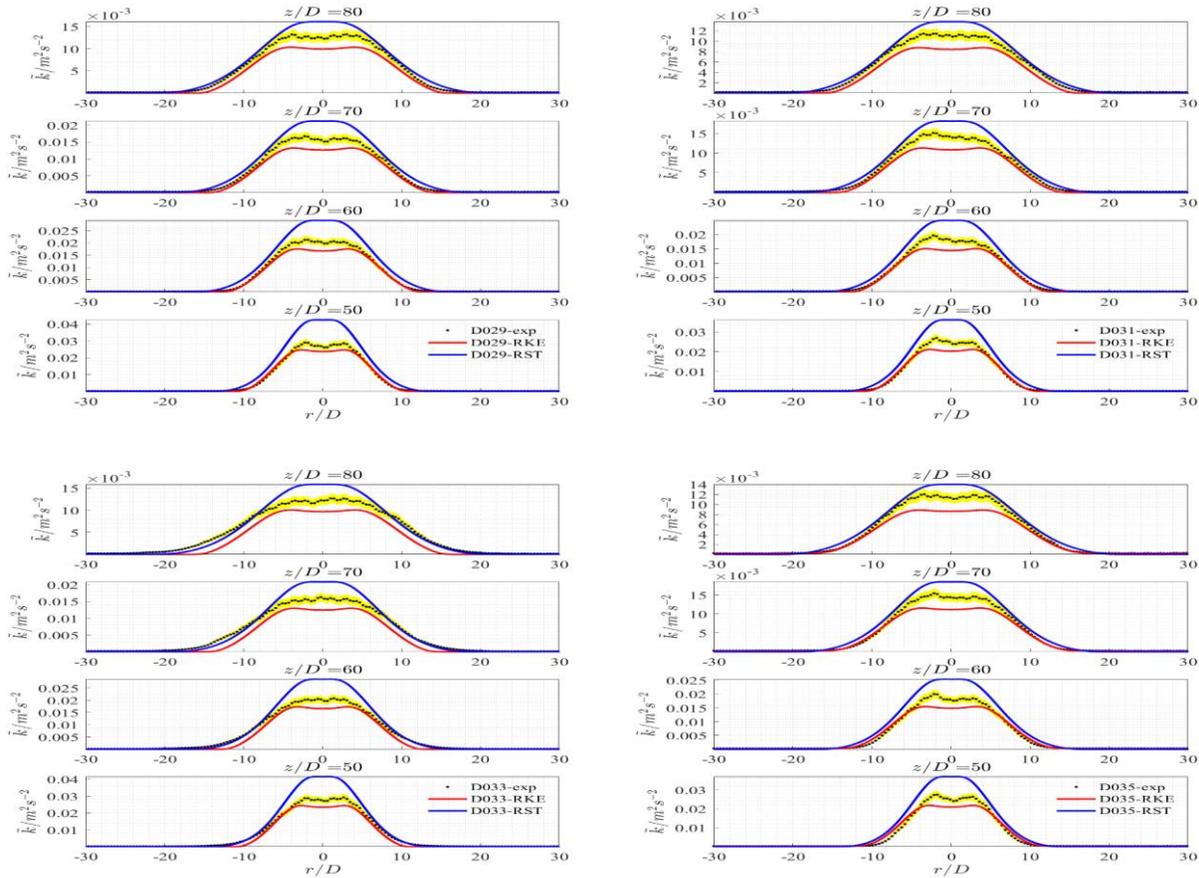

**Figure 8. Turbulent kinetic energy profiles**

### 4.3 Turbulent Kinetic Energy Production

In the previous section it has been demonstrated that the RKE model tend to underpredict the turbulent kinetic energy. In order to investigate the major reason for underprediction of the turbulent kinetic energy, it is necessary to have a deeper look at the production term of turbulent kinetic energy as well. As turbulent flows are always dissipative, production of turbulence is essential to sustain turbulence kinetic energy. The turbulent kinetic energy production term in Einstein notation, is expressed as:

$$P_k = -\overline{u_i'u_j'}\frac{\partial U_i}{\partial x_j} \quad (7)$$

For round jet in cylindrical coordinate, Eq. (7) can be rewritten as:

$$P_k = -\overline{u_z'u_r'}\left(\frac{\partial U_r}{\partial x_z}+\frac{\partial U_z}{\partial x_r}\right) - \overline{u_z'u_z'}\frac{\partial U_z}{\partial x_z} - \overline{u_r'u_r'}\frac{\partial U_r}{\partial x_r} \quad (8)$$

The comparison between simulation results and experiments for turbulence production is presented in Fig. 9. Double peaks in the radial profiles of $P_k$ are observed in both simulations and experiments. This is caused by the two shear layers present on both sides of the jet. It can be observed that for all the four runs, turbulence production is well predicted by both the RKE and RST models in the shear layer region, while in the jet center region the RST model over-estimates turbulence production, and the RKE model underestimates turbulence production. As the turbulent production given by the RKE model is always lower than experimental values, it is then understandable why the RKE model under-predicts turbulence kinetic energy profiles as inferred above. The good agreement of turbulence production in the center region given by the RST model indicates that the overestimation of turbulence kinetic energy in the center region is due to the underestimation of turbulence dissipation.

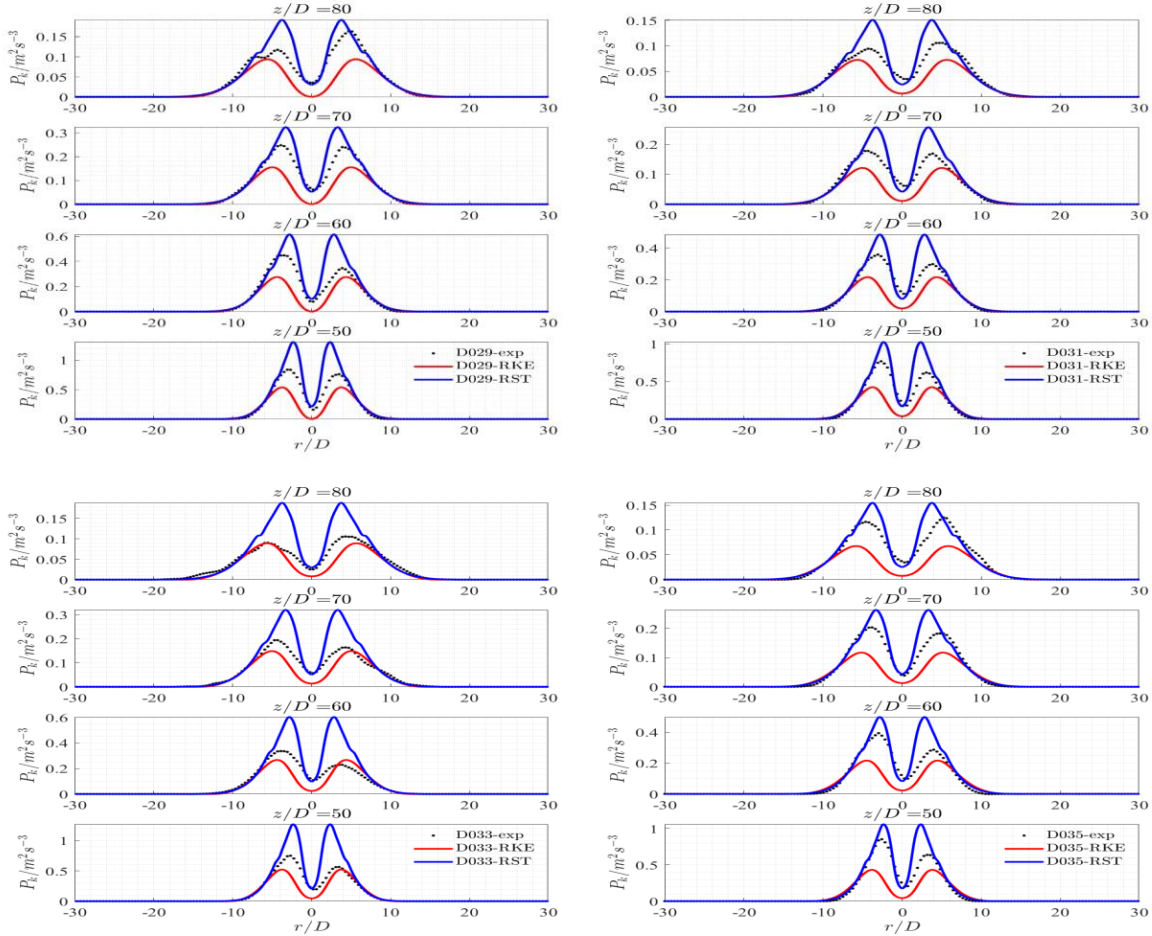

**Figure 9. Turbulent production profiles**

Since the production term can be expanded into three individual terms which are composed of various product of velocity gradient and Reynolds stresses, as shown in Eq. (8), the contribution from each term can be illustrated by plotting them separately (see Fig. 10, where the radial profiles of each term of $P_k$ are shown for all experimental runs at different axial locations). The circled numbers in the legend represent the contribution to $P_k$ in the same order as shown in Eq. (8). It can be seen that the product of the shear stress with the mean strain rate tensor is the main contributor to the turbulence production in the shear layer. In the jet center, instead, the main contributor is the product of the streamwise normal stress with the gradient of the mean streamwise velocity along the axial direction. For the RKE model, each dominant term has a smaller magnitude compared to experimental values, the total of which account for the underprediction of turbulence production. On the contrary, the RST model results in larger values for each dominant term, therefore yielding overprediction of turbulence production.

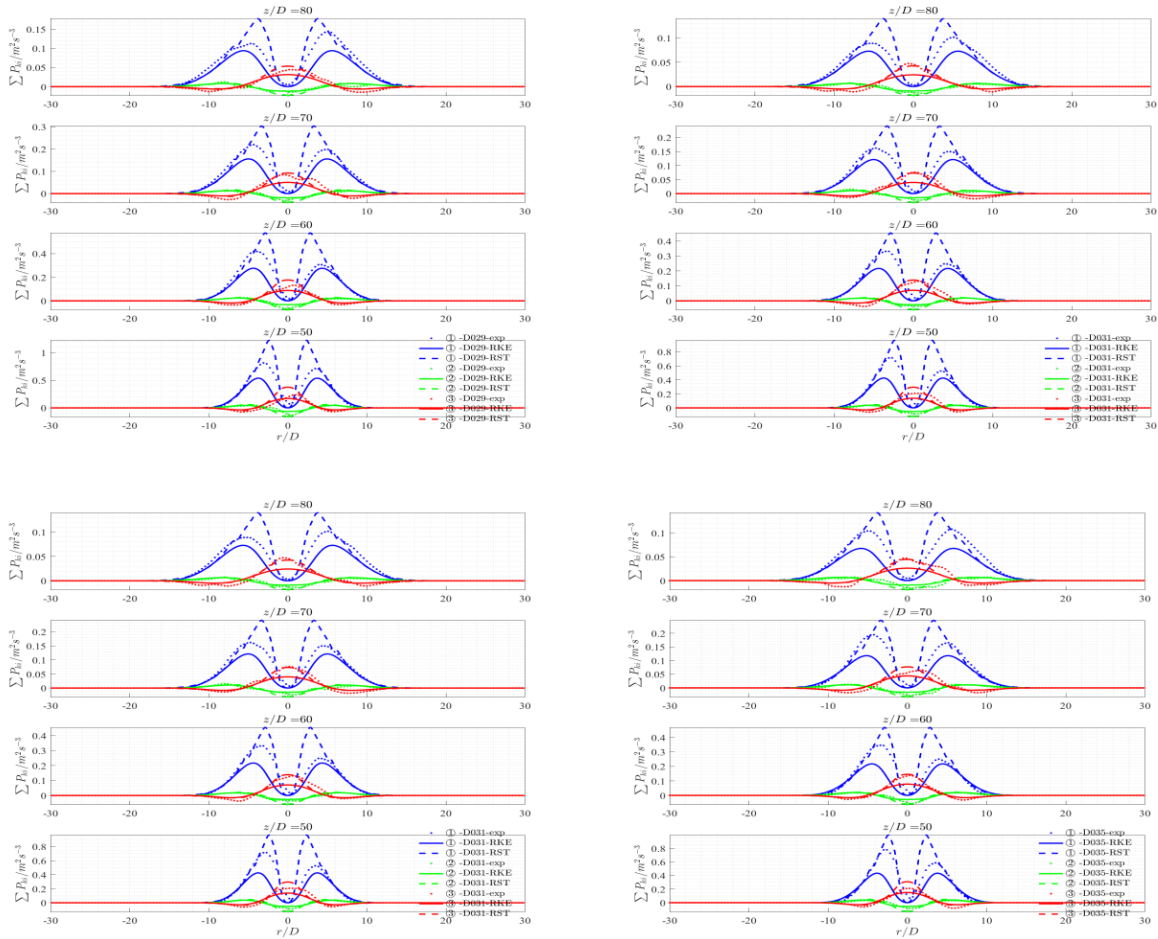

**Figure 10. Turbulent production budgets profiles**

### 4.4 Impact of RKE model parameters

The results discussed in the previous sections have demonstrated that the RKE model tends to underestimate turbulence production as well as turbulence turbulent kinetic energy in the jet self-similar region. To test whether a change in turbulence production and dissipation would result in improved result, we consider the turbulence dissipation transport equation:

$$\frac{\partial \rho \varepsilon}{\partial t} + \frac{\partial (\rho \varepsilon U_j)}{\partial x_j} = \frac{\partial}{\partial x_i}\left[\left(\mu + \frac{\mu_t}{\sigma_k}\right)\frac{\partial \varepsilon}{\partial x_j}\right] + C_{\varepsilon 1}\frac{\varepsilon}{k}(P_k + C_{\varepsilon 3}\frac{\mu_t}{\rho Pr_t}\frac{\partial \rho}{\partial x_j} \cdot g_i) - C_{\varepsilon 2}\rho\frac{\varepsilon^2}{k} \qquad (9)$$

where the default model coefficients $C_{\varepsilon 1}$, $C_{\varepsilon 2}$ and $C_{\varepsilon 3}$ are not universal since they have been tuned based on a limited set of flow conditions. Their values are critical for calculating the dissipation rate and therefore determine the accuracy of the RKE model. To investigate the influence of their values on the simulation results, case D029 was taken as reference and the values of $C_{\varepsilon 1}$, $C_{\varepsilon 2}$ where modified, while $C_{\varepsilon 3}$ was kept as default value since the results discussed in this paper have shown that the RKE model underpredicts turbulence kinetic energy regardless of density differences.

In Fig. 11 and Fig. 12, the streamwise velocity and turbulent kinetic energy profiles are reported for case D029, with varying model coefficients. It can be observed that $C_{\varepsilon 1}$ does not affect the mean velocity profiles while having a slight improvement on turbulent kinetic energy in the jet center. This can be expected from Eq. (9), since $C_{\varepsilon 1}$ is responsible for magnifying turbulence production. Turbulence production predicted by RKE is fairly small in the center region according to Fig. 8, which contributes less to the transport equation. As a result, $C_{\varepsilon 1}$ made a slight influence on reducing dissipation. $C_{\varepsilon 2}$ instead is decisive for accurate predictions. Compared to default value 1.90, slightly small $C_{\varepsilon 2}$ shows good agreement of velocity and turbulent kinetic energy profiles for all the locations. From this experience, it can be seen that the RKE model is very sensitive to $C_{\varepsilon 2}$ and the selection of its value is of importance. This conclusion is consistent with Lai et al. [7] and Thiesset et al. [23], who both agreed on reduced $C_{\varepsilon 2}$, even having a disagreement on the specific value.

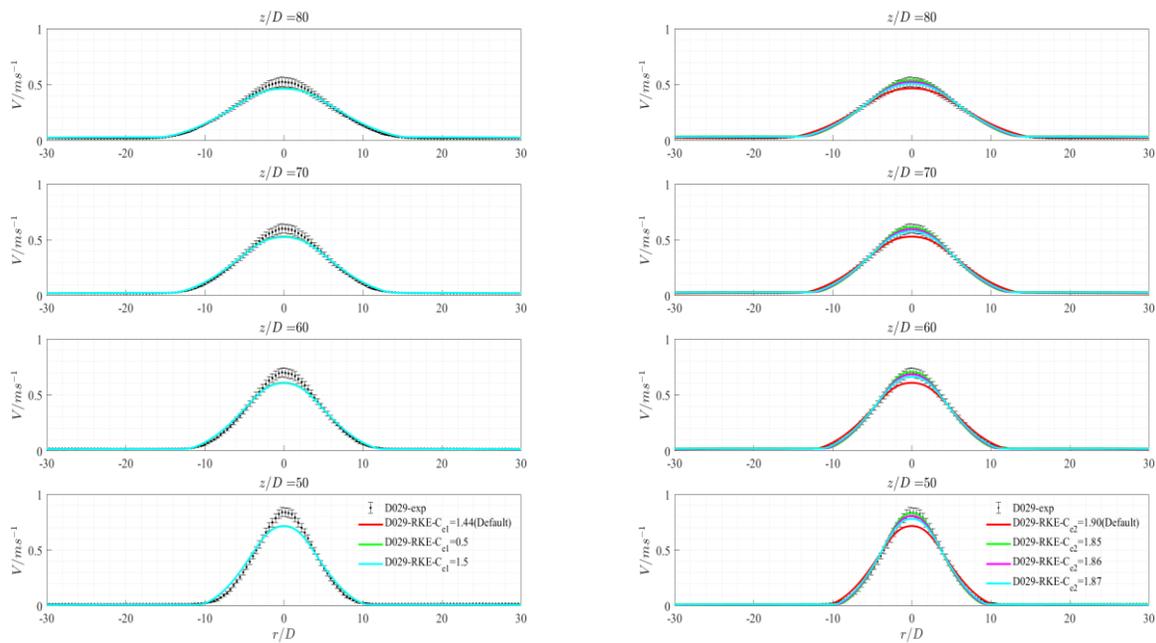

**Figure 11. Streamwise velocity profiles by modified model coefficients**

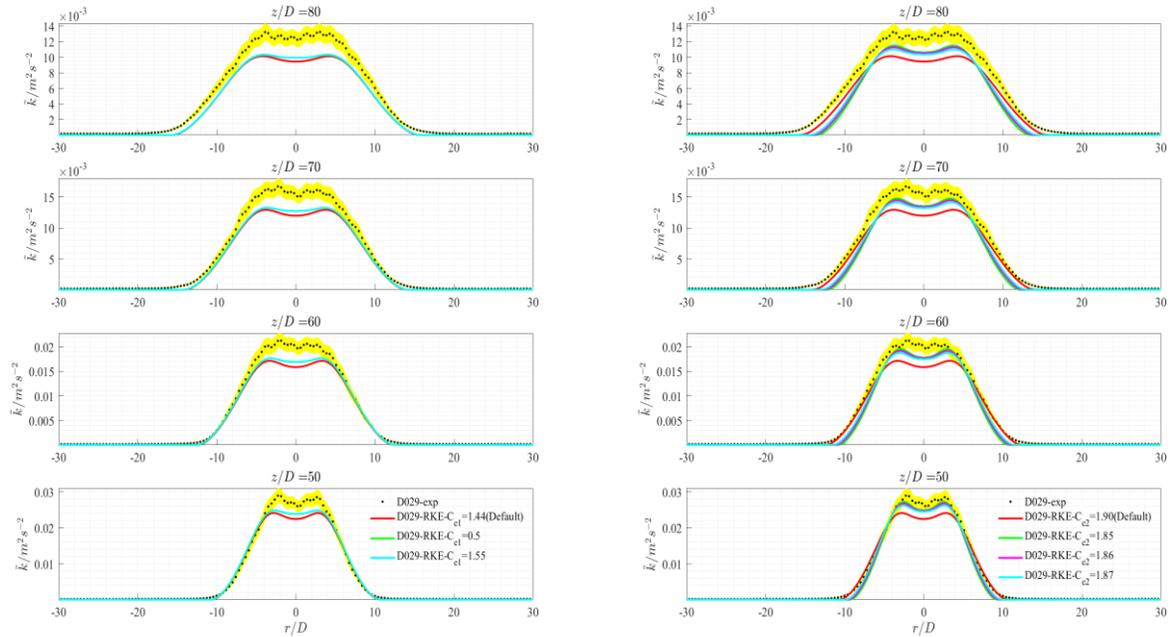

**Figure 12. Turbulent kinetic energy profiles by modified model coefficients**

## 5. CONCLUSION

The performance of the RKE model and the RST model has been assessed against high-resolution experimental data for turbulent round jets with and without density differences. It is observed that, although the default RKE model can predict well shear stresses, it tends to underestimate streamwise velocity and turbulence kinetic energy in the jet center region. A deeper analysis of the turbulence production term indicates that the RKE model also underestimates turbulence production. The RST model, instead, results in good agreement for mean streamwise velocity profiles and shear stress, while it overestimates turbulence kinetic energy in the center region. No difference in model performance for the turbulence quantities has been observed comparing results for uniform density and cases with density differences. This should be further investigated for cases with higher density differences when the buoyancy effect is even stronger. However, the streamwise velocity profiles are better predicted in cases where the light fluid is injected into heavy fluid, compared to the case in which heavy fluid is injected into the light fluid. In addition, the model coefficients of the turbulence dissipation equation for the RKE model were modified to investigate their effects on the simulation results. It was found that the RKE model is very sensitive to coefficient $C_{\varepsilon 2}$ and its value has a critical influence on the model accuracy.

## ACKNOWLEDGMENTS

Gratitude must be expressed for the instruction and help from colleagues in the ECMFL group of the University of Michigan and the generous support of Nuclear Energy University Program (Project no. 14-6552) of the U. S. Department of Energy.